# A Comparative Study of Various Distance Measures for Software fault prediction

Deepinder Kaur

*Assistant Professor, CSE Deptt. ChitkaraUniversity*

*Rajpura,Punja,India*

deepinder.kaur@chitkara.edu.in

*Abstract*— **Different distance measures have been used for efficiently predicting software faults at early stages of software development. One stereotyped approach for software fault prediction due to its computational efficiency is K-means clustering, which partitions the dataset into K number of clusters using any distance measure. Distance measures by using some metrics are used to extract similar data objects which help in developing efficient algorithms for clustering and classification. In this paper, we study K-means clustering with three different distance measures Euclidean, Sorensen and Canberra by using datasets that have been collected from NASA MDP (metrics data program) .Results are displayed with the help of ROC curve. The experimental results shows that K-means clustering with Sorensen distance is better than Euclidean distance and Canberra distance.**

## I. INTRODUCTION

Cluster analysis is the mission of assembling a set of objects in such a way that objects in the one group are more similar to each other than to those in other groups. Clustering algorithm use data metrics and apply various distance measures for assessing cluster distance. The distance measure plays an important role in obtaining correct clusters. Out of several clustering algorithms K-means clustering is considered to be efficient due to its ease of use. K-means clustering uses different distance measures for detecting more efficient results. The mail goal of K-means clustering algorithm is to cluster the software projects in fault free and fault prone groups. Fault prediction is to expect the possibility that the software contains faults or defects. Software defect can be requirement defect, design defect, code defect, test case defect and other product defect.[7] Several distance measures react differently for same algorithm. In this paper, we study three different distance measures Euclidean distance, Sorensen distance, Canberra distance and applied them on K-means clustering algorithm and evaluate the efficient distance measure. Datasets that have been used to evaluate results have been collected from NASA MDP (metrics data program) [1]. ROC curve i.e. receiver operator characteristic curve has been drawn to better predict the quality[6]. The experimental results demonstrate that Sorensen distance is most efficient amongst the three distance functions or measures.

## II. DISTANCE MEASURES

Different measures of distance or similarity are convenient for different types of analysis. For numeric datasets, often used distance functions are Euclidean distance, Manhattan distance[2], Sorensen or Bray Curtis distance, Canberra distance, Chebyshev distance .Similarly for Boolean datasets and other non numeric datasets other distance measures are used. Image distance is commonly used distance function for images and colours datasets .In the current paper, we study basic Euclidean distance, Sorensen or Bray Curtis distance and Canberra distance.

A **metric** on a set X is a function (called the *distance function* or simply **distance**)

$d : X \times X \to \mathbf{R}$ (where **R** is the set of real numbers). For all *x*, *y*, *z* in *X*, this function is required to satisfy the following conditions:

1. $d(x, y) \geq 0$ (*non-negativity*, or separation axiom)
2. $d(x, y) = 0$ if and only if $x = y$ (coincidence axiom)
3. $d(x, y) = d(y, x)$ (*symmetry*)
4. $d(x, z) \leq d(x, y) + d(y, z)$ (*Triangle inequality*).

Numeric distance measures:

i. **Euclidean Distance**

Euclidean distance computes the root of square difference between co-ordinates of pair of objects.

$$S_i^{(t)} = \left\{ x_j : \| x_j - m_i^{(t)} \| \leq \| x_j - m_{i^*}^{(t)} \| \text{ for all } i^* = 1, \dots, k \right\}$$

ii. **Manhattan Distance**

Manhattan distance computes the absolute differences between coordinates of pair of objects

$$S_i^{(t)} = \{ \|xi - yi\| \} \text{ for all } i = 1, \dots, k$$

iii. **Sorensen Distance**

Sorensen distance is a normalization method that views the space as grid similar to the city block distance. Sorensen distance has a nice property that if all coordinates is positive; its value is between zero and one. The normalization is done using absolute difference divided by the summation[3]

$$S_j^{(t)} = \left\{ x_j : \frac{\| x_j - m_j \|}{\| x_j + m_j \|} \text{ for all } j = 1, \dots, k \right\}$$

iv. **Canberra Distance**

Canberra distance examines the sum of series of a fraction differences between coordinates of a pair of objects. This distance is very sensitive to a small change when both coordinates are nearest to zero [6].

$$S_j^{(t)} = \left\{ x_j : \frac{\| x_j - m_j \|}{\| x_j \| + \| m_j \|} \text{ for all } j = 1, \ldots, k \right\}$$

### III. CLUSTERING

Clustering is a process of partitioning a set of data into a set of meaningful sub-classes, called clusters that help users to understand the natural grouping or structure in a data set. Seliya N. and Khoshgoftaar T.M. investigated semi supervised learning approach for classifying data to improve software quality rather than supervised and unsupervised learning only [6]. K-means clustering is one of the best examples of semi-supervised learning [8].

K-means is a clustering algorithm depends upon iterative location that partitions dataset into K no. of clusters by standard Euclidean distance.

A. **K-means clustering algorithm with Euclidean Distance**
Let X={$x_1,x_2...x_k$} be set of data and M={$m_1,m_2....m_k$}
1) Select a number (K) of cluster centers - centroids at random
2) Assign every item to its nearest cluster center using Euclidean distance

$$S_i^{(t)} = \left\{ x_j : \| x_j - m_i^{(t)} \| \leq \| x_j - m_{i^*}^{(t)} \| \text{ for all } i^* = 1, \ldots, k \right\}$$

3) Move each cluster center to the mean of its assigned items

$$m_i^{(t+1)} = \frac{1}{|S_i^{(t)}|} \sum_{x_j \in S_i^{(t)}} x_j$$

4) Repeat steps 2,3 until convergence or change in cluster assignment less than a threshold.

B. **K-means clustering algorithm with Canberra distance**
Let X={$x_1,x_2...x_k$} be set of data and M={$m_1,m_2....m_k$}
1) Select a number (K) of cluster centers - centroids at random
2) Assign every item to its nearest cluster center using Sorensen distance

$$S_j^{(t)} = \left\{ x_j : \frac{\| x_j - m_j \|}{\| x_j + m_j \|} \text{ for all } j = 1, \ldots, k \right\}$$

3) Move each cluster center to the mean of its assigned items

$$m_i^{(t+1)} = \frac{1}{|S_i^{(t)}|} \sum_{x_j \in S_i^{(t)}} x_j$$

4) Repeat steps 2,3 until convergence or change in cluster assignment less than a threshold.

C. **K-means clustering algorithm with Sorensen distance**
Let X={$x_1,x_2...x_k$} be set of data and M={$m_1,m_2....m_k$}
1) Select a number (K) of cluster centers - centroids at random.
2) Assign every item to its nearest cluster center using Canberra distance

$$S_j^{(t)} = \left\{ x_j : \frac{\| x_j - m_j \|}{\| x_j \| + \| m_j \|} \text{ for all } j = 1, \ldots, k \right\}$$

3) Move each cluster center to the mean of its assigned items

$$m_i^{(t+1)} = \frac{1}{|S_i^{(t)}|} \sum_{x_j \in S_i^{(t)}} x_j$$

4) Repeat steps 2, 3 until convergence or change in cluster assignment less than a threshold.

### IV. ROC CURVE

Receiver-operating characteristic (ROC) curves are an excellent way to compare diagnostic tests. The curve is created by plotting the true positive rate or probability of detection (PD) against the false positive rate or probability of false alarms (PF) at various threshold settings. The scales of the roc curve that is PD and PF are the basic measures of accuracy and are easily read from the plot.

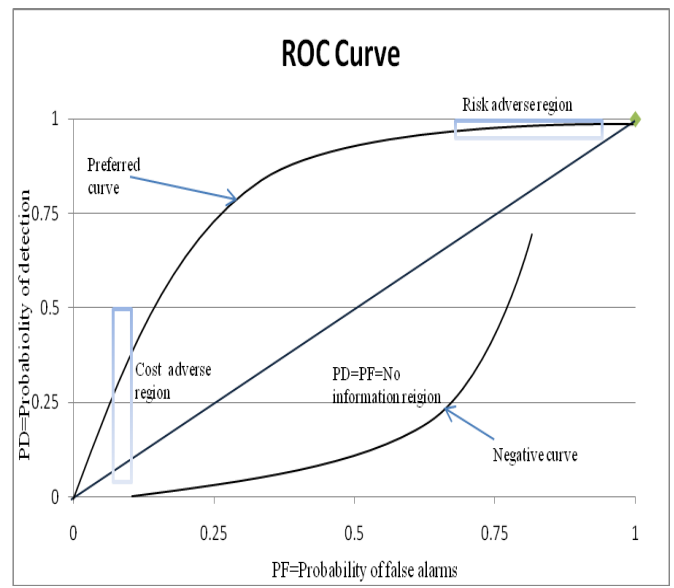

Fig.1 ROC Curve

Generally ROC curve has concave shape that starts at point (0, 0) and end at (1,1). High PD and high PF is beneficial for safety critical systems as faults identification is of more value than validating false alarms, This region is known as Risk adverse region .Similarly the region that defines low PD and low PF which is considered to be good for the organizations having limited Verification & Validation budgets is known as Cost adverse region. Sometimes low PD and high PF region or negative curve is preferred by some of the software projects. As the PD decreases and PF increases, the probability that modules can be classified incorrectly increases [6].

### V. RESULTS AND DISCUSSION

The datasets used for predicting defect prone modules using K-means clustering with efficient distance measures have been picked up from NASA metrics data program. Three projects CM1, PC1 and JM1 are used with Requirement, code and join metrics (obtained by natural join of requirement and code metrics). The results have been collected and shown in tables.([4],[8])

Table I K-means clustering algorithm with Euclidean distance –PC1

| Evaluation measures/Projects | PC1 | | |
|---|---|---|---|
| | Requirement | Code | Join |
| PD | 0 | 0 | 0.99219 |
| PF | 0 | 0 | 0.79578 |

TABLE II K-means clustering algorithm with Euclidean distance-CM1

| Evaluation measures/Projects | CM1 | | |
|---|---|---|---|
| | Requirement | Code | Join |
| PD | 0 | 0 | 1 |
| PF | 0 | 0 | 0.99729 |

TABLE III K-means clustering algorithm with Canberra distance-PC1

| Evaluation measures/Projects | PC1 | | |
|---|---|---|---|
| | Requirement | Code | Join |
| PD | 0.074766 | 0.97368 | 0.99748 |
| PF | 0.11737 | 0.94762 | 0.71304 |

TABLE IV K-means clustering algorithm with Canberra distance-CM1

| Evaluation measures/Projects | CM1 | | |
|---|---|---|---|
| | Requirement | Code | Join |
| PD | 0.10145 | 0.99795 | 1 |
| PF | 0.15 | 0.95721 | 0.96718 |

TABLE V K-means clustering algorithm with Sorensen distance-PC1

| Evaluation measures/Projects | PC1 | | |
|---|---|---|---|
| | Requirement | Code | Join |
| PD | 0.76471 | 0.57945 | 1 |
| PF | 0.9 | 0.37033 | 0.98925 |

TABLE VI K-means clustering algorithm with Sorensen distance-CM1

| Evaluation measures/Projects | CM1 | | |
|---|---|---|---|
| | Requirement | Code | Join |
| PD | 0.53623 | 0.83333 | 0.98795 |
| PF | 0.6 | 0.4442 | 0.96175 |

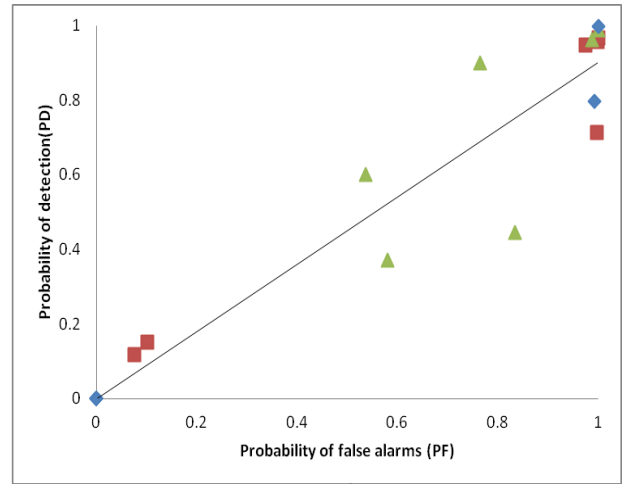

Fig. 3 Resulted ROC curve

In ROC Curve the results of K-means clustering algorithm with Euclidean distance are shown with blue colour diamonds and the results of K-means clustering algorithm with Canberra distance are shown with red colour squares and the results of K-means clustering algorithm with Sorensen distance are shown with green colour triangles. The results shows that if we use Euclidean distance measure then it gives best performance in case of join metrics as the values of PD and PF are high thus it can be use for projects having high risk ,all other metrics results in 0,0 lies on no information region. In case of Canberra distance measure, it is efficient for low budget projects as its requirement and code metrics PD and PF values lie near to cost adverse region, But in case of join metrics it is worse than Euclidean distance measure for some projects. In case of Sorensen distance the results with each metric is more accurate that give better fault prediction for high risk projects as well as low budget projects than Euclidean distance and Canberra distance measure as shown in Fig. 3. K-means clustering with Sorensen distance will thus more accurately cluster modules into fault vulnerable and non-fault vulnerable as compared to other two distance measures, it has high probability of detection (PD) and less probability of false alarms (PF).

VI. CONCLUSION AND FUTURE SCOPE

In this paper we compared the results of K-means clustering algorithm with three different distance measures that is basic Euclidean distance, Canberra distance and Sorensen distance to forecast the fault vulnerability at premature phase of software life process along with available data that may help the software practitioners to erect more accurate projects [5]. The results with Sorensen distance are more accurate in case of high risk projects. The results with Canberra distance are efficient for projects having low verification and validation budgets. The results with Euclidean distance are good in some projects. It is intelligible from ROC curve. From the results the software practitioners should make more effort for projects which are fault vulnerable to make them fault free. This is beneficial for improving software reliability. In addition comparison of other algorithms with different distance measures can be done to achieve high quality software fault predictors.